\documentclass[%
 reprint,
superscriptaddress,
 aps,
floatfix
]{revtex4-2}

\usepackage{graphicx}
\usepackage{xcolor}
\usepackage{dcolumn}
\usepackage{bm}
\usepackage{comment}
\usepackage{wrapfig}
\usepackage{natbib}
\usepackage{multirow}
\usepackage{makecell}
\usepackage{float}
\usepackage[section]{placeins}
\usepackage{amsmath}
 \usepackage{amsfonts}
 
\usepackage{units}
\usepackage{color}
\usepackage{url}

\usepackage[colorlinks]{hyperref}
\hypersetup{%
	plainpages=true,
	breaklinks=true,
	hypertexnames=false,
	pageanchor=true,
	colorlinks=true,
	linkcolor={blue},
	citecolor={red},
	urlcolor={blue},
	anchorcolor={black}
}

\newcommand{\ket}[1]{|#1\rangle}

\begin{document}

\preprint{APS/123-QED}

\title{A quantum thermal machine surpassing the classical thermodynamic limit on precision}

\author{Simon Sundelin}
\email{simsunde@chalmers.se}
\affiliation{Department of Microtechnology and Nanoscience,
Chalmers University of Technology,
412 96 Gothenburg, Sweden}

\author{Ludvig Nordqvist}
\affiliation{Department of Microtechnology and Nanoscience,
Chalmers University of Technology,
412 96 Gothenburg, Sweden}

\author{Khalak Mahadeviya}
\affiliation{School of Physics, Trinity College Dublin, College Green, Dublin 2, D02 K8N4, Ireland}

\author{Vyom Kulkarni}
\affiliation{Department of Microtechnology and Nanoscience,
Chalmers University of Technology,
412 96 Gothenburg, Sweden}

\author{Mohammed Ali Aamir}
\affiliation{Department of Microtechnology and Nanoscience,
Chalmers University of Technology,
412 96 Gothenburg, Sweden}

\author{Mark T. Mitchison}
\email{mark.mitchison@kcl.ac.uk}
\affiliation{Department of Physics, King’s College London, Strand, London, WC2R 2LS, United Kingdom}
\affiliation{School of Physics, Trinity College Dublin, College Green, Dublin 2, D02 K8N4, Ireland}

\author{Simone Gasparinetti}
\email{simoneg@chalmers.se}
\affiliation{Department of Microtechnology and Nanoscience,
Chalmers University of Technology,
412 96 Gothenburg, Sweden}

\date{\today}

\begin{abstract}

Precision in nonequilibrium processes comes at a thermodynamic cost: suppressing fluctuations generally requires increased dissipation. Thermodynamic uncertainty relations (TURs) make this trade-off quantitative by linking current fluctuations to entropy production in classical stochastic dynamics. In the decade since its discovery, the canonical steady-state TUR and its finite-time generalizations have become a cornerstone of non-equilibrium thermodynamics, constraining the performance of molecular machines and allowing heat dissipation to be inferred from observable fluctuations. Whether the canonical TUR can be violated in a controlled quantum device remains an outstanding experimental question, in part because doing so requires resolving extremely small steady-state currents as well as their fluctuations. Here we experimentally show that steady-state quantum transport can surpass the precision permitted by the canonical TUR. We observe this violation in a superconducting quantum thermal machine coupled to a microwave waveguide acting as a cold bath and to a classical noise source providing an effective infinite-temperature bath. We observe a TUR ratio $\mathcal{Q} = 1.71 \pm 0.17$, in violation of the classical bound $\mathcal{Q}\geq 2$. Our results demonstrate a fundamental distinction between classical and quantum thermodynamics, paving the way for quantum thermal devices that achieve enhanced precision at reduced energy cost. 

\end{abstract}

\maketitle

Fluctuations cannot be neglected in small nonequilibrium systems, where matter and energy are exchanged in discrete quanta and transport is intrinsically noisy.  Thermodynamic uncertainty relations (TURs) bound the relative fluctuations of a current by the entropy production required to sustain it, such that precision cannot be increased above a certain limit without additional heat dissipation~\cite{barato_thermodynamic_2015,gingrich_dissipation_2016,pietzonka_finite-time_2017,horowitz_proof_2017, Timpanaro2019, Hasegawa2019_xft, koyuk_thermodynamic_2020}. 
This trade-off between fluctuations and dissipation has emerged as a foundational principle in the thermodynamics of far-from-equilibrium systems, shaping the performance of biomolecular clocks~\cite{Barato2016} and sensors~\cite{harvey_universal_2023}, while introducing precision as a new figure of merit that competes on the same footing as the power and efficiency of nanoscale heat engines~\cite{pietzonka_universal_2018}.
Beyond their fundamental significance, TURs have unlocked the practical ability to infer the heat dissipated to hidden environmental degrees of freedom via experimentally accessible observables~\cite{Hwang2018, Seifert2019, Martinez2019, Manikandan2020,Skinner2021}.

For classical nonequilibrium steady states, the canonical TUR~\cite{barato_thermodynamic_2015,gingrich_dissipation_2016} reads
\begin{equation}
\label{eq:TURratio}
    \mathcal{Q} := \frac{\sigma}{k_B}\frac{D}{J^2} \geq 2,
\end{equation}
where $\sigma$ is the entropy production rate, $J$ is the average current flowing through the system, $D$ is the diffusion coefficient characterizing the current's fluctuations, and $k_B$ is the Boltzmann constant, with the combination $\mathcal{Q}$ referred to as the TUR ratio. However, as quantum mechanics sets in at small length and temperature scales, the assumptions underpinning Eq.~\eqref{eq:TURratio}---namely, classical Markovian dynamics obeying local detailed balance---break down. Indeed, theoretical modelling has predicted that current fluctuations in open quantum systems can violate the canonical TUR~\cite{Brandner2018, ptaszynski_coherence-enhanced_2018, agarwalla_assessing_2018, liu_thermodynamic_2019, guarnieri_thermodynamics_2019, kalaee_violating_2021, van_vu_thermodynamics_2022}. These predictions raise the tantalising prospect of quantum devices performing tasks such as thermoelectric energy conversion~\cite{ptaszynski_coherence-enhanced_2018} or timekeeping~\cite{meier_precision_2025} with higher precision than would be possible classically with the same resources. Such applications require steady-state operation to produce a sustained, useful output. Yet observing a quantum enhancement of thermodynamic precision in a nonequilibrium steady state remains an outstanding problem. A major challenge is to measure heat dissipation and current fluctuations in an open quantum system without measurement backaction destroying its underlying coherent dynamics. 
This difficulty has so far restricted experimental investigations of TURs in quantum-mechanical systems to transient unitary dynamics~\cite{pal_experimental_2020} or to regimes where classical bounds are satisfied~\cite{yang_phonon_2020, Friedman2020, Barker2025}.

\begin{figure*}[t]
    \includegraphics[width=17.5cm]{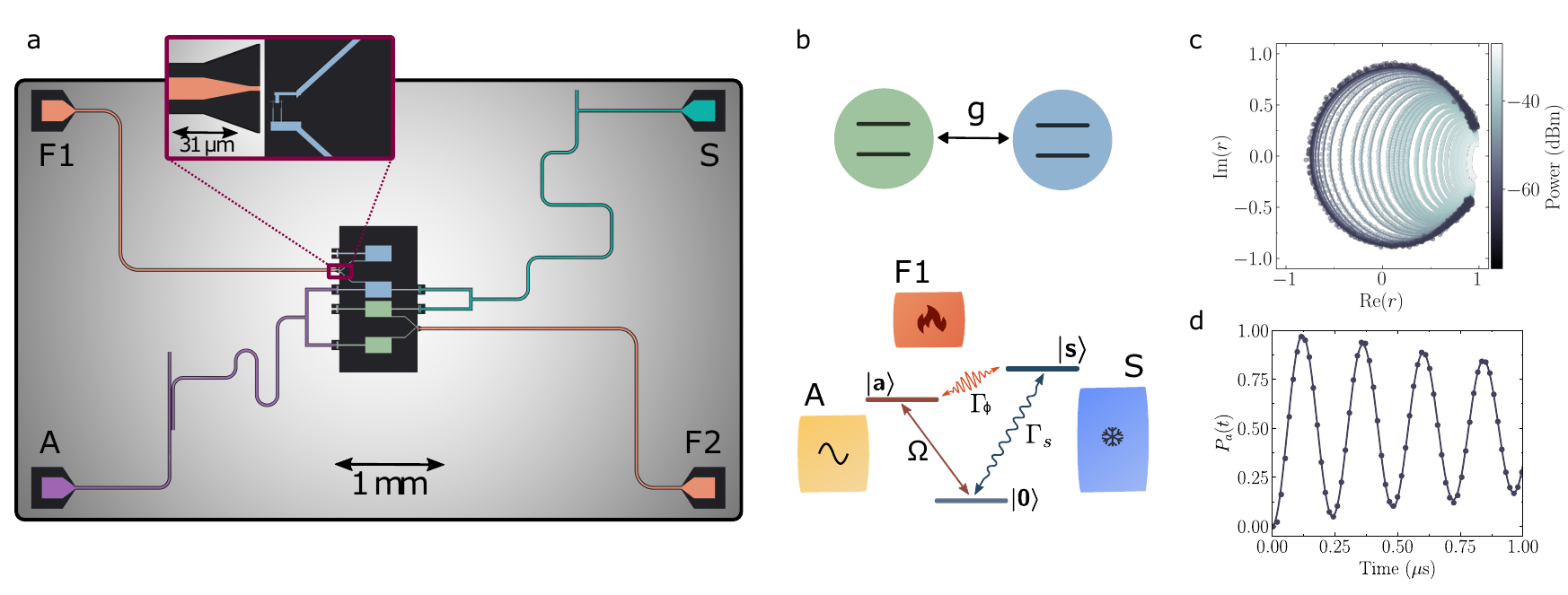}
    \caption{\label{fig1} Device architecture and operating principle. (a) Rendered false-colour micrograph of the superconducting artificial molecule, composed of two flux-tunable transmons coupled to symmetry-selective microwave waveguides S and A. The flux lines used to inject longitudinal noise are shown in orange. Inset: superconducting quantum interference device (SQUID) of one transmon. (b) Hybridized level structure and transport cycle of the artificial molecule. \(|0\rangle\), \(|a\rangle\), and \(|s\rangle\) denote the ground, antisymmetric, and symmetric molecular eigenstates, respectively. The upper-state splitting is \(2g\), where \(g\) is the coherent inter-qubit coupling strength. A coherent drive with Rabi frequency \(\Omega\) excites the \(|0\rangle \leftrightarrow |a\rangle\) transition. Longitudinally coupled noise induces incoherent transfer between \(|a\rangle\) and \(|s\rangle\) at an equal rate \(\Gamma_{\Phi}\) back and forth, thereby acting as an effective infinite-temperature bath. The symmetric state \(|s\rangle\) decays into the cold waveguide at a rate \(\Gamma_s\), completing the transport cycle. (c)~Reflection coefficient of waveguide~S as a function of probe frequency and power. Fits to a model based on input-output theory and Lindblad master equation yield the relevant coupling rates and confirm that the symmetric waveguide is the dominant radiative channel. (d) Rabi oscillations in the population of $|a\rangle$, measured under resonant driving of the \(|0\rangle \leftrightarrow |a\rangle\) transition through the A-resonator, with the solid line showing a fit to the time-domain response. }
\end{figure*}

Here we experimentally demonstrate a steady-state violation of the canonical TUR in a superconducting quantum thermal machine coupled to a microwave waveguide.
Using near-quantum-limited heterodyne detection of the microwave field emitted into the waveguide, we resolve the mean photon current and its fluctuations in the sub-attowatt regime. 
This unprecedented resolution is enabled by a novel approach to infer microwave intensity fluctuations---nominally defined by a more challenging photon-counting measurement~\cite{landi_current_2024}---from the accessible heterodyne signal.
Combined with spectroscopic measurements of the microwave temperature~\cite{scigliuzzo_primary_2020}, our technique allows us to assess the precision and entropy production of our device purely in terms of its outputs, without invasive measurements of its quantum state.  
We observe relative current fluctuations below the limit set by the TUR on classical non-equilibrium steady states with the same rate of entropy production. Observing this violation provides a thermodynamic witness of nonclassical transport; at the same time, it opens a path towards exploiting precision enhancements in quantum thermal devices. 

\section*{Superconducting thermal machine}
Superconducting circuits combine engineered dissipation, tunable coupling to propagating microwave modes, and precise control of nonequilibrium driving, making them an attractive platform to investigate fundamental questions in quantum thermodynamics~\cite{pekola_towards_2015,pekola_colloquium_2021,aamir_engineering_2022,ronzani_tunable_2018,gubaydullin_photonic_2022,maillet_electric_2020,aamir_thermally_2025,sundelin_quantum_2026}. 
Our device [Fig.~\ref{fig1}(a)] consists of two strongly coupled, flux-tunable transmon qubits forming an artificial molecule. We tune the transmons into resonance, so that the single-excitation manifold of the molecule is described by the hybridized symmetric and antisymmetric modes
$|s\rangle=(|10\rangle+|01\rangle)/\sqrt{2}$ and $|a\rangle=(|10\rangle-|01\rangle)/\sqrt{2}$, where $\ket{ij}$ denotes the joint state of the first qubit in state $\ket{i}$ and the second qubit in state $\ket{j}$~[Fig.~\ref{fig1}(b)]. At this operating point, we measure mode frequencies $\omega_s/2\pi = 5.9$ GHz and $\omega_a/2\pi = 5.5$ GHz, respectively, corresponding to a splitting $2g/2\pi = 400$ MHz, where $g$ is the inter-qubit coupling strength.

The molecule is coupled to two microwave waveguides, denoted S and A, via two coplanar waveguide resonators with fundamental frequencies $\omega_f/2\pi = 6.0$ GHz and $\omega_r/2\pi = 7.1$ GHz and linewidths $\kappa_f/2\pi = 39$ MHz and $\kappa_r/2\pi = 0.5$ MHz, respectively. Both resonators act as Purcell filters, inhibiting or enhancing the decay of the molecular transitions into the waveguides~\cite{houck_controlling_2008}. The resonators are coupled to multiple points of the molecule, leveraging the spatial symmetry of the molecular eigenstates to selectively enhance or suppress coupling via electromagnetic interference, which provides an additional layer of selectivity~\cite{aamir_engineering_2022}. The S-resonator is designed to enhance the emission of $|s\rangle$ into waveguide S, while the A-resonator enables dispersive readout of the molecular states while protecting them from decay.
Finally, each qubit is coupled to a flux line (F1 and F2), which we use to tune the qubit frequency as well as to supply finite-frequency longitudinal noise, as explained below.
We perform our measurements in a dilution refrigerator at a base temperature below 10 mK unless specified otherwise.

\begin{figure}
    \includegraphics[width=8.5cm]{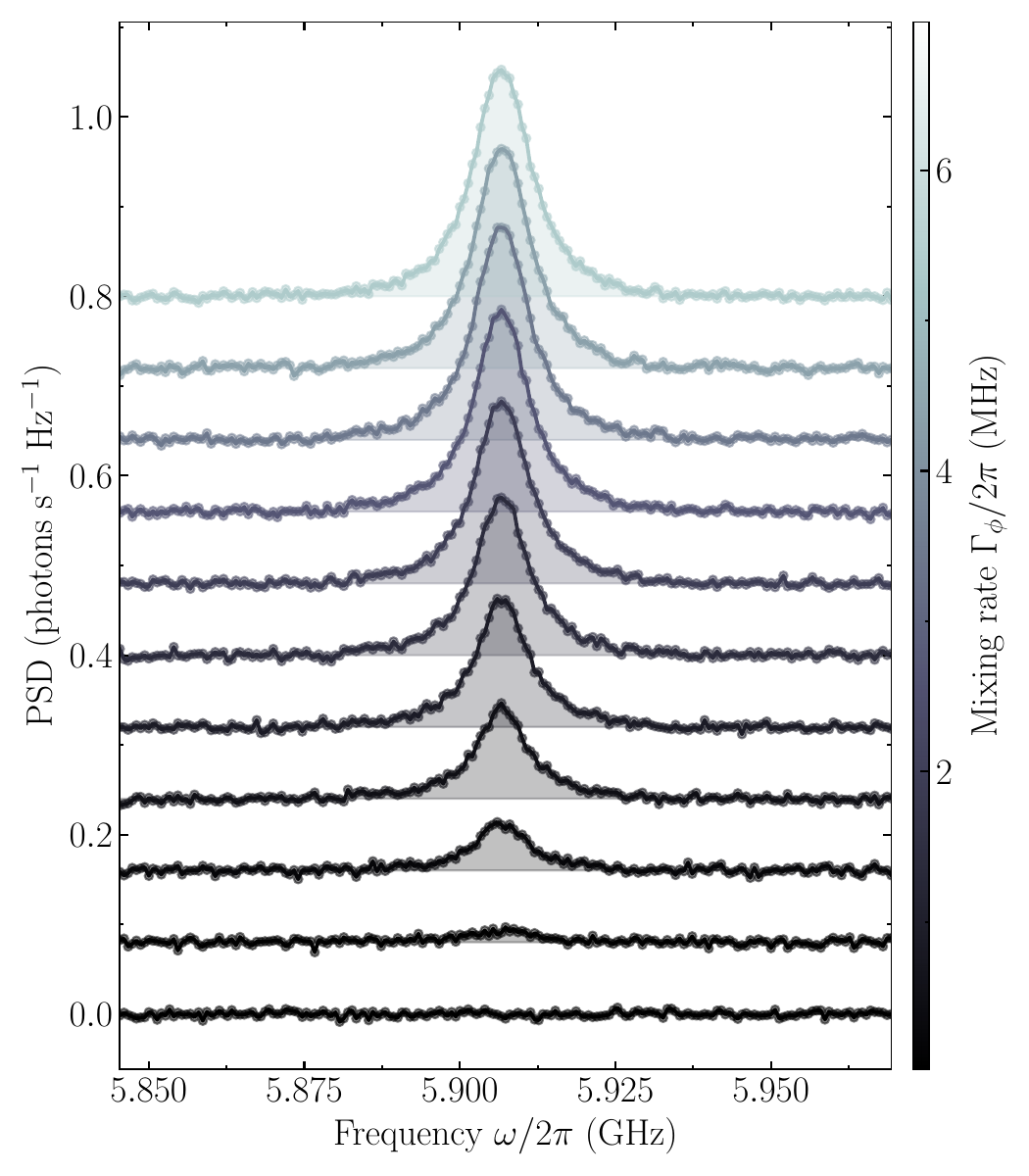}
    \caption{\label{fig2} Power spectral density of the output field in the symmetric waveguide for increasing mixing rates at a fixed drive rate. The growing emission peak signals heat transferred into the cold bath, and its integrated area yields the emitted heat current.}
\end{figure}

We operate the device as a quantum thermal machine [Fig.~\ref{fig1}(b), bottom]. The thermal cycle goes as follows.
We coherently drive the $\ket{0} \leftrightarrow \ket{a}$ transition via the A-resonator. By injecting broadband classical noise to F1 at frequencies encompassing $2g$, we induce stochastic transitions between states $\ket{s}$ and $\ket{a}$ with equal rate~\cite{sundelin_quantum_2026}. Finally, state $\ket{s}$ spontaneously decays into waveguide S, bringing the system back to the ground state. In this regime, the device functions as a driven thermal accelerator; waveguide A acts as a coherent work repository, flux line F1 as an effective infinite-temperature bath, and waveguide S as a cold bath. This system is closely related to the three-level transport model studied theoretically by Kalaee \emph{et al.}~\cite{kalaee_violating_2021}.

We characterize the coupling of $\ket{s}$ to waveguide S by measuring the reflection coefficient of S as a function of frequency and drive power~[Fig.~\ref{fig1}(c)]. Fitting a model based on input--output theory and Lindblad master equation to the data, we extract a linewidth of $\Gamma_s/2\pi = 7.2$ MHz. By contrast, $\ket{a}$ is strongly protected against emission and remains long lived. We verify coherent control of $\ket{a}$ by measuring Rabi oscillations on the lower transition $\ket{0} \leftrightarrow \ket{a}$ [Fig.~\ref{fig1}(d)]. From a $T_1$ measurement, we also measure its decay rate $\Gamma_a/2\pi = 30$ kHz.

We next characterize the noise-activated heat transport through the machine. We apply filtered white noise to the flux line of one qubit. The noise has flat spectral density $S_{\Phi}(\omega)$ centered at the frequency of the level splitting, $2g$, and a bandwidth of 50~MHz. In the molecular energy eigenbasis, its dominant effect is to induce bidirectional $|a\rangle \leftrightarrow |s\rangle$ transitions, effectively acting as an infinite-temperature bath in the $\{|a\rangle, |s\rangle\}$ subspace~\cite{sundelin_quantum_2026}. The strength of this coupling is controlled by the applied noise amplitude through the spectral density at the transition frequency, $\Gamma_{\Phi} \propto S_{\Phi}(2g)$.

We observe the resulting transport in the power spectral density (PSD) of the outgoing microwave field (Fig.~\ref{fig2}). In the absence of injected noise, we observe no measurable emission in the symmetric waveguide, whereas at finite noise power a clear emission peak emerges whose integrated area yields the emitted photon flux. We calibrate the PSD independently using the resonance-fluorescence spectrum of the strongly driven symmetric transition (see Methods). In the fully resolved Mollow-triplet regime, the measured spectrum is compared with the corresponding theoretical lineshape, which determines the gain of the measurement chain and thereby places the measured signal on an absolute scale of emitted microwave power. Independently, we extract the incoherent transfer rate $\Gamma_{\Phi}$ from Rabi measurements of the antisymmetric mode at increasing noise powers.

\section*{Thermodynamic uncertainty relation}
To test the thermodynamic uncertainty relation, we evaluate the steady-state TUR ratio defined in Eq.~\eqref{eq:TURratio}. In our experiment, 
$J$ is the mean photon current emitted into the symmetric waveguide, $D$ is its diffusion coefficient, $\sigma = \hbar\omega_s J/T$ is the entropy-production rate, and $T$ is the temperature of the electromagnetic modes in the waveguide~S.  In the operating regime studied here, $\sigma$ is dominated by heat emitted into waveguide S in the form of photons with energy $\hbar\omega_s$, because (i) emission into waveguide A is strongly suppressed, (ii) absorption from waveguide S is negligible at low enough temperature, and (iii) the large coherent tone sustaining the transport cycle acts as a nearly ideal work repository. (For a theoretical discussion regarding this last point, see Ref.~\cite{janovitch_bridging_2026}.) The temperature of the cold waveguide is determined independently from spectroscopic calibration measurements described in Methods and, together with the measured current, determines the corresponding entropy-production rate.

\begin{figure}
    \includegraphics[width=8.5cm]{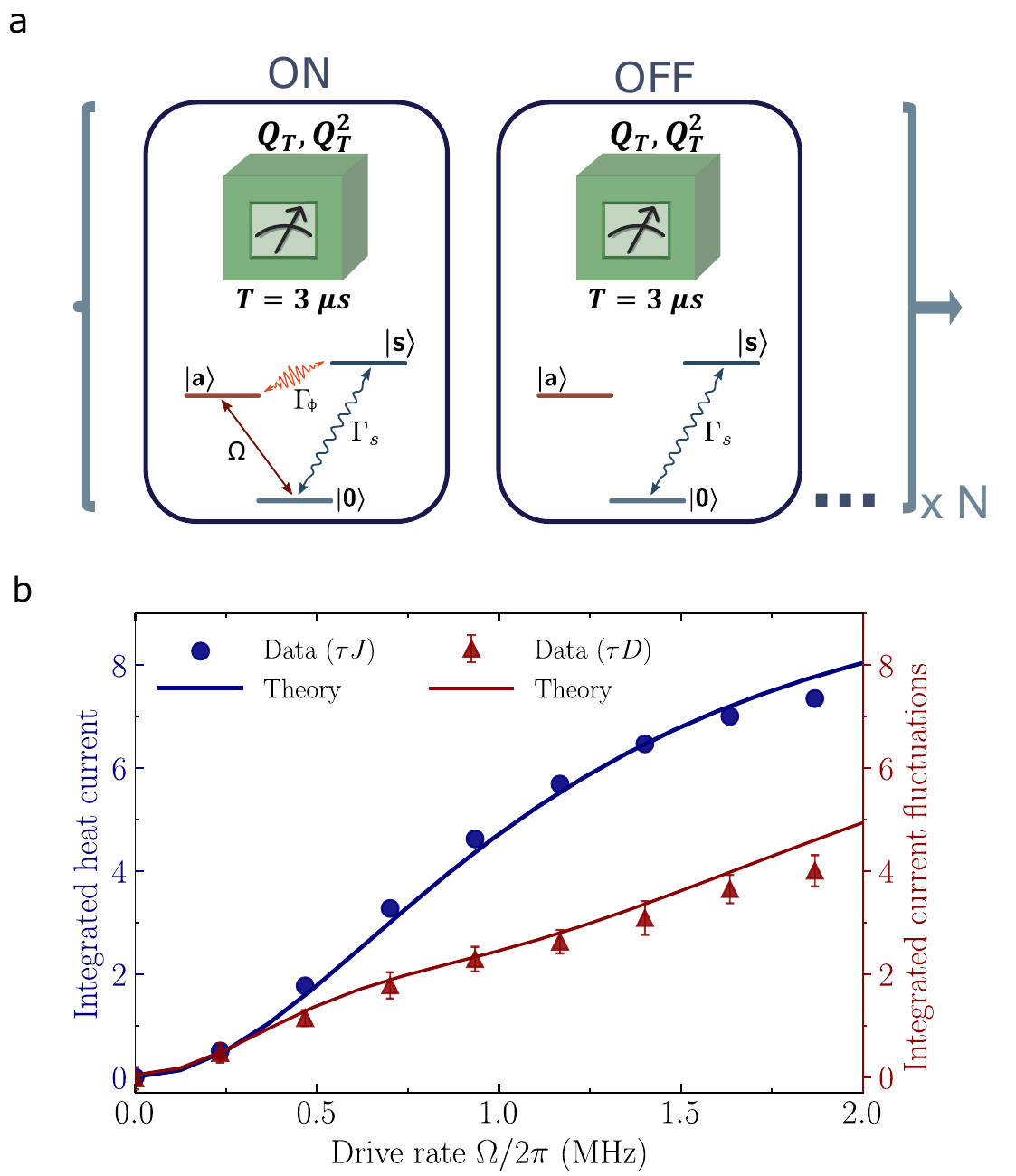}
    \caption{\label{fig3} Extraction of photon-current moments from continuous microwave measurements. (a) Interleaved ON/OFF protocol used to determine the first and second moments of the integrated heterodyne power. In the ON state, the coherent drive and injected flux noise sustain transport through the machine, whereas the OFF state provides the background reference of the measurement chain. (b) Measured integrated current $\tau J$ (circles) and current fluctuations $\tau D$ (triangles) as functions of drive strength over the integration time \(\tau=3\,\mu{\rm s}\) and a fixed mixing rate $\Gamma_{\Phi}/2\pi=3.4$ MHz. Solid lines show full counting statistics predictions evaluated using independently characterized device parameters and no free parameters.}
\end{figure}

The mean current is formally defined by $J = \langle N_\tau\rangle/\tau$ and its diffusion coefficient is $D = {\rm Var}(N_\tau)/\tau$, where $N_\tau$ is the number of photons emitted into the symmetric waveguide over a large timespan $\tau$~\cite{landi_current_2024}. However, a direct measurement of $N$ would require continuous detection of itinerant microwave photons, a yet-unresolved experimental challenge~\cite{bozyigit_antibunching_2011,eichler_experimental_2011,virally_discrete_2016}. Therefore, we develop an alternative characterization of photon-current noise using heterodyne measurements of the microwave field amplitude. We analyze the integrated heterodyne power over a measurement window \(\tau\),
\begin{equation}
    \hat{Q}_{\tau} = \int_0^{\tau} dt\, \bar{S}^{\dagger}(t)\bar{S}(t),
    \label{eq:Qtau}
\end{equation}
where \(\bar{S}(t)\) is the filtered output field containing both the emitted signal and the background noise of the measurement chain. The current is measured using an interleaved ON/OFF protocol [Fig.~\ref{fig3}(a)]. In the ON state, we simultaneously apply the coherent drive and injected mixing noise, while in the OFF state, we turn them both off. Measurements taken in the OFF state thus provide us with a reference containing only the background noise and the noise added by our amplification chain.

Using input--output theory, we relate the first moment of \(\hat{Q}_{\tau}\) to the emitted heat current \(J\) as (see Methods)
\begin{equation}
    \frac{\langle \hat{Q}_{\tau}\rangle}{\tau} = J + \eta^{-1}\Lambda,
    \label{eq:meanQ}
\end{equation}
where $\eta$ is the measurement efficiency and \(\Lambda\) is the effective detection bandwidth set by the filter. Taking the difference between ON and OFF measurements eliminates the background contribution and yields
\begin{equation}
    J = \frac{1}{\tau}\left[\langle \hat{Q}_{\tau}\rangle_{\rm on} - \langle \hat{Q}_{\tau}\rangle_{\rm off}\right].
    \label{eq:Jextract}
\end{equation}
By performing a similar calculation for the second moment of \(\hat{Q}_{\tau}\), we obtain the diffusion coefficient
\begin{equation}
    D = \frac{1}{\tau}\left[{\rm Var}(\hat{Q}_{\tau})_{\rm on} - {\rm Var}(\hat{Q}_{\tau})_{\rm off}\right] - \frac{2-\eta}{\eta}J,
    \label{eq:Dextract}
\end{equation}
where the final term accounts for excess fluctuations in the ON-state heterodyne power that are unavoidably added by the amplification chain. To evaluate this, we need the measurement efficiency $\eta$ which can be estimated directly from the OFF-state cumulants as $\eta = \frac{2\langle \hat{Q}_T\rangle_{\rm off}}{3 {\rm Var}(\hat{Q}_T)_{\rm off}}\simeq 30\%$. See Methods for a discussion and the Supplementary Information for a full derivation.

\begin{figure}
    \includegraphics[width=8.5cm]{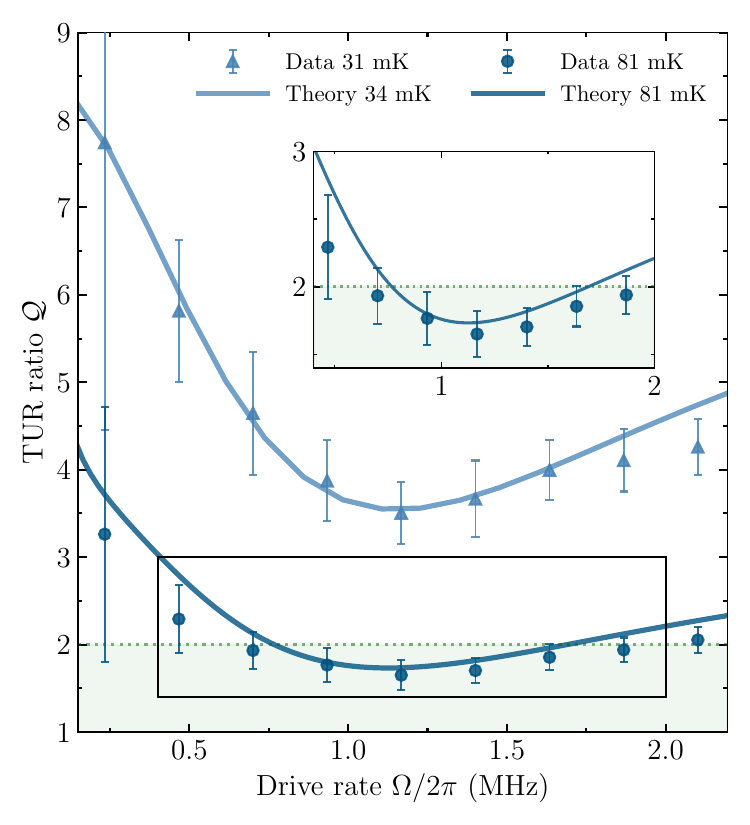}
    \caption{\label{fig4} Violation of the classical thermodynamic uncertainty relation. Measured TUR ratio $\mathcal{Q}=(\sigma/k_B)D/J^2$ as a function of drive strength for cold-bath temperatures of 31 mK and 81 mK. The dashed line marks the classical bound $\mathcal{Q}=2$. At 81 mK, the measured ratio falls below the classical threshold and enters the sub-classical regime predicted by full counting statistics (see inset for the same data on a larger scale). Solid lines show theoretical results evaluated using independently characterized device parameters, and error bars denote two standard deviations.}
\end{figure}

Using more than \(10^8\) repeated measurements in total, and an integration time of \(\tau = 3\,\mu{\rm s}\), we extract both \(J\) and \(D\) for different drive strengths and fixed mixing rate [Fig.~\ref{fig3}(b)].
We simulate the outcome of our measurements using
full-counting-statistics
and independently characterized device parameters (see Methods for details of the calculation), and find an excellent agreement between theory and experiment, without adjustable parameters.

With the mean current, diffusion coefficient, and cold-bath temperature in hand, we evaluate the TUR ratio according to Eq.~\eqref{eq:TURratio}. For classical nonequilibrium steady-state transport, the TUR reads \(\mathcal{Q}\geq 2\). Figure~\ref{fig4} shows the measured ratio as a function of drive strength \(\Omega_a\) for a fixed mixing rate \(\Gamma_{\Phi}/2\pi = 3.4\,\)MHz and for two values of the cold-bath temperature. The error bars in Fig.~\ref{fig4} represent two standard deviations and are dominated by sampling uncertainty. At the base temperature of 31\,mK, the extracted TUR ratio remains above the classical threshold and is highly sensitive to the bath temperature. At the elevated temperature of 81\,mK, however, the ratio reaches a minimum value of $\mathcal{Q}=1.71 \pm 0.17$. Within our experimental uncertainties, this demonstrates an experimental violation of the steady-state TUR in a driven superconducting quantum thermal machine.

\section*{Discussion}

The TUR was originally derived for classical, Markovian, stationary processes driven by thermal noise on a discrete state space~\cite{gingrich_dissipation_2016}, and later extended to overdamped Langevin dynamics of continuous variables~\cite{Dechant2018}.
Deviations from these assumptions can lead to violations even in the classical regime, e.g., feedback control~\cite{Barker2025} or mechanical inertia in continous-variable systems~\cite{Pietzonka2022}. However, our system has a discrete set of states and is well described (in a rotating frame) by autonomous Markovian dynamics obeying local detailed balance (see Methods). We therefore attribute the observed TUR violation to a dissipative quantum effect, which arises from the simultaneous contribution of quantum-coherent and entropy-producing processes to the transport cycle~\cite{kalaee_violating_2021}. Our results can also be understood as arising from a suppression of photon-current noise by the optical nonlinearity of the transmon, since linear scattering of noninteracting bosons cannot generate TUR violations~\cite{Saryal2019}. 

In contrast to other quantum effects, which are most pronounced at the lowest temperatures,
we find that the quantum enhancement of precision only occurs at intermediate temperatures. For the same drive and dephasing conditions as in the experiment, our theoretical model predicts an enhancement over a temperature range from approximately 60 mK to 180 mK. Higher temperatures would suppress quantum coherence, whereas the classical bound is always satisfied at lower temperatures because the entropy production entering the TUR ratio~\eqref{eq:TURratio} diverges. This trade-off underscores the interplay between quantum and thermal effects underlying the observed phenomenon.

In conclusion, we have experimentally demonstrated that steady-state quantum transport can surpass the precision allowed by the canonical TUR for classical nonequilibrium dynamics.  The key experimental advance that enabled our work is the direct extraction of both the mean current and its fluctuations from continuous microwave measurements in the sub-attowatt regime. This method enables the precision–dissipation trade-off to be evaluated from the experimentally accessible features of our device's output field, providing a ``black-box'' test of its thermodynamic characteristics. Our technique to measure current noise also opens the door to further experimental studies of other far-from-equilibrium fluctuation-dissipation relations~\cite{hasegawa_quantum_2020, tesser_out-of-equilibrium_2024, palmqvist_combining_2025, Brandner2025, friedman_thermodynamic_2020} in superconducting circuit platforms.

Our experiment shows that quantum coherence can act as an operational thermodynamic resource, shaping not only the states and spectra of a nanoscale device but also reducing the noise in its output. This demonstrates a profound difference between quantum and classical stochastic thermodynamics, challenging the expectation that quantum fluctuations always lead to increased uncertainty at the nanoscale. By placing quantum violations of classical TURs on a firm experimental footing, we take a concrete step towards future applications that exploit this effect for improved thermodynamic efficiency, such as autonomous quantum clocks whose precision can grow exponentially with entropy production~\cite{meier_precision_2025}.

\begin{acknowledgments}
The presented device design was assisted by the Python package QuCAT \cite{gely2020b} and was fabricated in Myfab Chalmers, a nanofabrication laboratory. This work received support from the Swedish Research Council via Grant No. 2021-05624, the Knut and Alice Wallenberg Foundation through the Wallenberg Center for Quantum Technology (WACQT), from the European Research Council via Grant No. 101041744 ESQuAT. M.T.M. is supported by a Royal Society University Research Fellowship. This project is co-funded by the European Union (Quantum Flagship project ASPECTS, Grant Agreement No. 101080167) and UK Research and Innovation (UKRI). Views and opinions expressed are, however, those of the authors only and do not necessarily reflect those of the European Union, Research Executive Agency or UKRI. 
\end{acknowledgments}

\newpage


\section*{Methods}

\subsection{Experimental setup}
We mount the device at the mixing-chamber stage of a dilution refrigerator and access its symmetric and antisymmetric modes through coupled coaxial lines. Attenuators at successive cryogenic stages strongly attenuate and thermalize the input signals, while circulators and cryogenic and room-temperature amplifiers route and amplify the output signals. The symmetric output line additionally includes a traveling-wave parametric amplifier (AI-TWPA-C, Arctic Instruments). A cryogenic bias tee combines the RF mixing-noise signal with the DC flux-bias current before delivering the combined signal to one of the qubit flux lines. We perform continuous-wave reflection spectroscopy using a vector network analyzer and acquire frequency-resolved spectra and power measurements using an IQ-based microwave transceiver (Quantum Machines OPX+). To generate broadband flux noise, we synthesize voltage noise with a bandwidth of 50~MHz and upconvert it to a center frequency near the molecular splitting \(2g\) before injecting it through the flux line. To control the base temperature during the thermometry calibration, we apply a current to a resistor anchored to the refrigerator base plate. The Supplementary Material provides the wiring diagram and further experimental details.

\subsection{Device design and coupling rates}

The device consists of two nominally identical, flux-tunable transmon qubits that are capacitively coupled to form a hybridized artificial molecule with symmetry-selective coupling to two microwave environments. Each transmon comprises capacitor pads shunted by a SQUID, which allows its transition frequency to be tuned by current applied through a nearby flux line. At the operating point used in the experiment, the molecule is described by the bare Hamiltonian
\begin{equation}
H_0=\sum_{j=1}^2 \hbar\omega_j b_j^{\dagger}b_j + \hbar g\left(b_1^{\dagger}b_2+b_2^{\dagger}b_1\right)+\sum_{j=1}^2\frac{\hbar\alpha_j}{2}b_j^{\dagger}b_j^{\dagger}b_j b_j,
\label{eq:SI_device_hamiltonian_bare}
\end{equation}
where $b_j$ is the annihilation operator of transmon $j$, $\omega_j$ are the bare transition frequencies, $\alpha_j$ are the anharmonicities, and $g$ is the inter-qubit coupling. When the two transmons are tuned to resonance, the single-excitation states $|10\rangle$ and $|01\rangle$ hybridize into the symmetric and antisymmetric molecular states
\begin{equation}
|s\rangle=\frac{|10\rangle+|01\rangle}{\sqrt{2}},
\qquad
|a\rangle=\frac{|10\rangle-|01\rangle}{\sqrt{2}},
\end{equation}
separated by an energy $2g$. 

The hybridized states are distinguished by their exchange symmetry and therefore couple differently to the surrounding microwave structure \cite{aamir_engineering_2022}. Transitions that preserve the molecular symmetry couple primarily to waveguide S, whereas symmetry-changing transitions couple primarily to waveguide A. In practice, this selectivity is reinforced by coupling the molecule to two different resonators. On the antisymmetric side, a readout resonator is used both for spectroscopy and for delivering the coherent drive. On the symmetric side, a filter resonator with a designed linewidth $24$ MHz, enhances the coupling of the symmetric transition to waveguide S while suppressing the density of states as seen by other transitions, thereby making state $|s\rangle$ competitively short lived compared to all other states in the molecule.

The coupling rate of the symmetric mode to the output waveguide is extracted from reflection spectroscopy on the $|0\rangle\leftrightarrow|s\rangle$ transition, with a reflection coefficient modelled as \cite{lu_characterizing_2021}
\begin{equation}
r\left(\omega-\omega_s\right)=1-\frac{i \Gamma_s \Gamma_{1}\left(\omega-\omega_s-i \Gamma_{2s}\right)}{\Omega_s^2 \Gamma_{2s}+\Gamma_{1s}\left[\left(\omega-\omega_s\right)^2+\Gamma_{2s}^2\right]}.
\end{equation}
Here, $\Gamma_s$ denotes the coupling rate between the symmetric mode and its waveguide while $\Gamma_s^\prime$ accounts for all other decay channels, so that $\Gamma_{1s}=\Gamma_s+\Gamma_s^\prime$ and $\Gamma_{2s}=(\Gamma_s+\Gamma_s^\prime)/2+\gamma_{s\phi}$, where $\gamma_{s\phi}$ is the pure-dephasing rate. The Rabi frequency is given by $\Omega_s=2\sqrt{P_{\rm in}A\Gamma_s/\hbar\omega_s}$, where $P_{\rm in}$ is the input power and $A$ is the total attenuation of the input line. Fitting the complex reflection coefficient [Fig. \ref{fig1} (b)] as a function of probe detuning and input power yields $\Gamma_s$ and a dephasing rate of $\gamma_{s \phi} = 490$ kHz. Owing to the strong separation of timescales between the long-lived antisymmetric state and the rapidly decaying symmetric state, the two modes may be regarded as effectively isolated. This approximation is supported by the nearly three-orders-of-magnitude separation between their decay rates, suggesting that residual nonradiative or cross-coupled loss channels are at most comparable to the weak decay of the antisymmetric mode and therefore negligible relative to the dominant symmetric radiative channel. The antisymmetric mode correspondingly retains a coherence time of $T_2 \approx 1.2~\mu$s.

\subsection{Cold-bath temperature calibration}

\begin{figure}[t]
    \includegraphics[width=8.5cm]{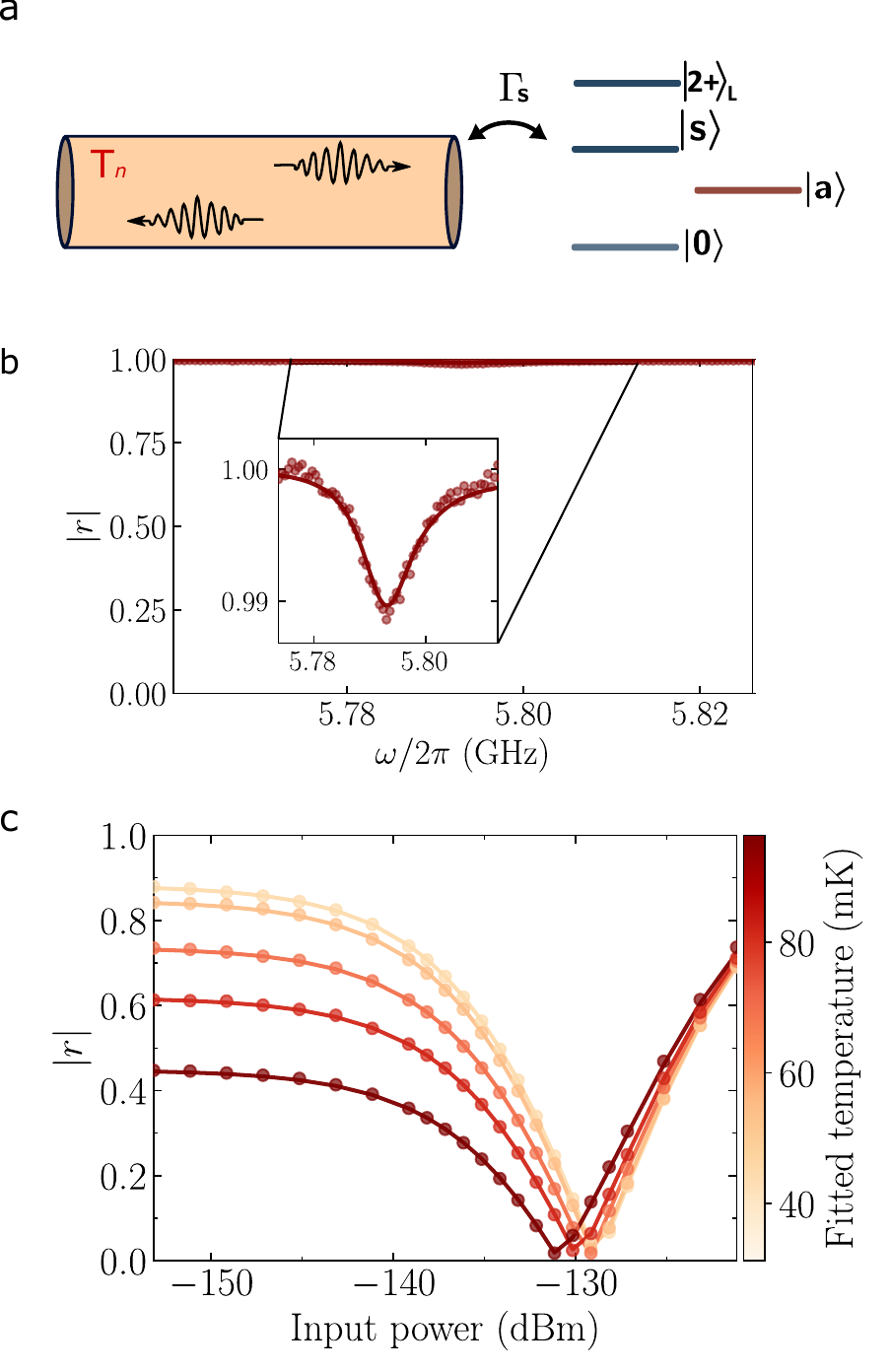}
    \caption{\label{meth1}Cold-bath temperature calibration based on a four-level molecular spectroscopy model. (a) Schematic of the four-level model. The waveguide at temperature \(T_n\) couples radiatively, with rate \(\Gamma_S\), to the symmetric transitions \(|0\rangle\leftrightarrow|s\rangle\) and \(|s\rangle\leftrightarrow|2+\rangle_L\), while the antisymmetric state \(|a\rangle\) remains dark to the waveguide but is weakly coupled to an environment at a similar temperature. Finite temperature produces thermal population in both \(|s\rangle\) and \(|a\rangle\). (b) Method A: reflection spectroscopy of the \(|s\rangle\leftrightarrow|2+\rangle_L\) transition at elevated temperature, obtained by sweeping the probe frequency across the transition. The solid line is a fit to the four-level model used to determine the effective waveguide temperature (see Supplementary information for details). (c) Method B: low-power on-resonance reflection measurements of the \(|0\rangle\leftrightarrow|s\rangle\) transition at \(\omega_d=\omega_s\), acquired across the experimental temperature range. The solid lines show fits to the calibrated reflection model used to extract the cold-bath temperature at different base-plate temperatures of the cryostat.}
    
\end{figure}

The on-resonance reflection amplitude for a qudit coupled to a waveguide depends directly on the thermal occupation of the incoming field, so that the photon population can be inferred from a calibrated scattering measurement at low enough power~\cite{scigliuzzo_primary_2020}. In our flux-tunable device, the same low-power response is also affected by residual dephasing and other nonthermal broadening channels. We therefore combine two complementary calibration methods. Method A determines the thermal occupation at an elevated reference temperature by sweeping the probe frequency across the second symmetric transition ($|s\rangle\leftrightarrow|2+\rangle_L$) of the four-level molecular spectrum [Fig.~\ref{meth1}(a,b)]. At this known temperature, Method B uses the on-resonance reflection response of the first symmetric transition ($|0\rangle\leftrightarrow|s\rangle$) to determine the nonthermal response parameters, which are subsequently held fixed when extracting the temperature at other heating settings [Fig.~\ref{meth1}(c)].

We first use Method A to calibrate the elevated-temperature occupation by sweeping the probe frequency across the \(|s\rangle\leftrightarrow|2+\rangle_L\) transition and fitting the resulting reflection spectrum [Fig.~\ref{meth1}(b)]. In this fit, the thermal occupation \(n\), drive rate, and relative dipole matrix element of the second transition are left as free parameters, while the radiative coupling rates \(\Gamma_s\) and \(\Gamma_a\) are fixed to values independently extracted from spectroscopy of the first symmetric and antisymmetric transitions, respectively. This gives a fitted waveguide temperature of \(96.3\)~mK at a base-plate heating power of 200~\(\mu\)W. This method is robust against residual pure dephasing. At a representative point in the experimentally relevant parameter regime, the change in the reflection magnitude produced by a \(10\%\) increase in thermal occupation is more than 600 times larger than that produced by the same relative increase in the pure-dephasing rate. For further details, see Supplemental information.

Having established this reference temperature, we apply Method B to a low-power, on-resonance reflection measurement of the \(|0\rangle\leftrightarrow|s\rangle\) transition recorded at the same heating power. With the thermal occupation fixed to the value obtained using Method A, this measurement determines the residual nonthermal response parameters of the thermometer. We then keep these parameters fixed and fit on-resonance reflection measurements acquired at other heating settings, using the thermal occupation \(n\) as the only free parameter [Fig.~\ref{meth1}(c)]. Although Method B is more sensitive to dephasing, anchoring it to Method A thereby provides a calibrated thermometer for the waveguide modes over the temperature range relevant to the TUR analysis.

\subsection{Extraction of mean currents and diffusion coefficient from heterodyne power}

To extract the mean photon current and its fluctuations from continuous heterodyne measurements, we model the detected signal using standard input-output theory in the presence of
amplifier noise. The output field from the device is written as the sum of the incoming field and the emitted system field, while the amplification chain adds a statistically independent bosonic noise mode. We assume the latter has stationary white-noise statistics with occupation $n_h$, corresponding to a measurement efficiency $\eta = \frac{1}{1+n_h} < 1$. In the large-gain limit, the amplified signal can be written as~\cite{clerk_introduction_2010}
\begin{equation}
S(t)=b_{\mathrm{o}}(t)+h^{\dagger}(t),
\end{equation}
where $b_{\mathrm{o}}(t)$ and $h(t)$ are bosonic ladder operators representing the output field and the added noise of the amplification chain, respectively. The measured heterodyne record is digitally filtered around the emission band, giving a finite-bandwidth signal \(\bar{S}(t)\), and converted into an integrated dimensionless power observable

\begin{equation}
\label{eq:SI_power}\hat{Q}_\tau=\int_0^\tau dt\,\bar{S}^{\dagger}(t)\bar{S}(t).
\end{equation}

Under the large-bandwidth assumption relevant to the experiment, the mean integrated power takes the form
\begin{equation}
\frac{\langle \hat{Q}_\tau\rangle}{\tau}=J+ \frac{\Lambda}{\eta},
\label{eq:SI_mean_general}
\end{equation}
where $J$ is the mean emitted photon current, $\Lambda$ is the effective detection bandwidth, and we assume $\Lambda\tau \gg 1$. The first term in Eq.~\eqref{eq:SI_mean_general} is the emitted photon current while the second term represents the contribution from the background noise field $h(t)$. By interleaving ON and OFF measurements, where the OFF state contains only the measurement background (i.e., $J=0$), this offset is removed and the mean current is obtained as
\begin{equation}
J=\frac{1}{\tau}\left[\langle \hat{Q}_\tau\rangle_{\rm on}-\langle \hat{Q}_\tau\rangle_{\rm off}\right].
\label{eq:SI_J_onoff}
\end{equation}
The mean entropy production rate is then given by $\hbar\omega_s J/T$, because each photon dissipates a mean heat $\int{\rm d}\omega \,P(\omega) \hbar\omega \approx \hbar\omega_s$ into the microwave environment at temperature $T$~\cite{Landi2021}, where $P(\omega)$ is the power spectral density as plotted in Fig.~\ref{fig2}. 

The variance of $\hat{Q}_\tau$ similarly contains both the intrinsic current fluctuations of the system and the excess fluctuations introduced by the finite-efficiency detection chain. Under the same assumptions of large measurement bandwidth and integration time, and assuming that the noise field $h(t)$ has Gaussian statistics, we obtain
\begin{equation}
    \label{eq:SI_var_general}
    \frac{{\rm Var}(\hat{Q}_\tau)}{\tau}= D + \frac{2-\eta}{\eta} J + \frac{3\Lambda}{2\eta^2},
\end{equation}
where $D$ is the diffusion coefficient of the emitted photon current. The final term above represents a pure white-noise contribution to the heterodyne power fluctuations. Its specific prefactor $3/2$ follows from the boxcar filter used in the analysis; a similar expression with a different prefactor would be obtained for another choice of filter (see the Supplemental Information). To remove this white-noise contribution, we subtract the measured variances in ON/OFF configurations to find
\begin{equation}
\frac{1}{\tau}\left({\rm Var}[\hat{Q}_\tau]_{\rm on}-{\rm Var}[\hat{Q}_\tau]_{\rm off}\right)=D+\frac{2-\eta}{\eta}J.
\label{eq:SI_boxcar_final}
\end{equation}
The second term above arises as a consequence of mixing between the emitted field and the background noise (e.g.~correlators of the form  $\langle b^\dagger_{\rm o}(t) h(t) b_{\rm o}(t') h^\dagger(t')\rangle $). In the ideal limit of unit efficiency, $\eta\to 1$, this mixing increases the measured current fluctuations by an amount equal to the mean current $J$; equivalently, it increases the effective Fano factor~\cite{landi_current_2024} by 1. This can be understood as the minimum amount of intensity noise added by a phase-insensitive linear amplification chain, which would be absent if the intensity measurement were instead realized via direct detection of emitted photons~\cite{landi_current_2024}.

In our experiment, the integrated power is evaluated over a measurement window of $\tau=3$ $\mu$s using a digital boxcar filter with a theoretical bandwidth of $32.5$ MHz. The effective detection bandwidth entering the analysis is determined jointly by the digital boxcar filter and the frequency response of the full measurement chain. We assume that this bandwidth is sufficiently large that the filter memory time $\Lambda^{-1}$ is short compared with the timescales over which the first- and second-order field correlations vary ($g^{(1)}$ and $g^{(2)}$). We further assume the long-time limit $\Lambda \tau \gg 1$, with $\tau$ also large compared with the correlation time of $g^{(2)}(t)$. These assumptions are necessary to ensure the validity of Eqs.~\eqref{eq:SI_mean_general} and~\eqref{eq:SI_var_general}. The measurement efficiency can then be estimated directly from the ratio of these two equations in the OFF state, which yields
\begin{equation}
\eta \approx \frac{2\langle \hat{Q}_\tau\rangle_{\rm off}}{3 {\rm Var}[\hat{Q}_\tau]_{\rm off}}.
\label{eq:eff}
\end{equation}
These expressions directly underlie the estimators used in the main text. Full mathematical derivations including finite-bandwidth and finite-time corrections are given in the Supplementary Information.

\subsection{Calibration of power and TUR extraction}

We determine the absolute scaling between the measured heterodyne power spectral density and the emitted
power at the device output from a calibration based on the Mollow triplet of a driven two-level transition. In this calibration, we coherently drive the symmetric transition and record the output spectrum under conditions in which the radiative linewidth, thermal occupation and pure dephasing are independently known from spectroscopy. We fit the incoherent part of the measured spectrum using the well known Lindblad-master-equation and quantum-regression-theorem treatment of resonance fluorescence~\cite{carmichael_open_1993,bozyigit_antibunching_2011}; the full model is given in the Supplementary Information.

\begin{figure}[t]
    \includegraphics[width=8.5cm]{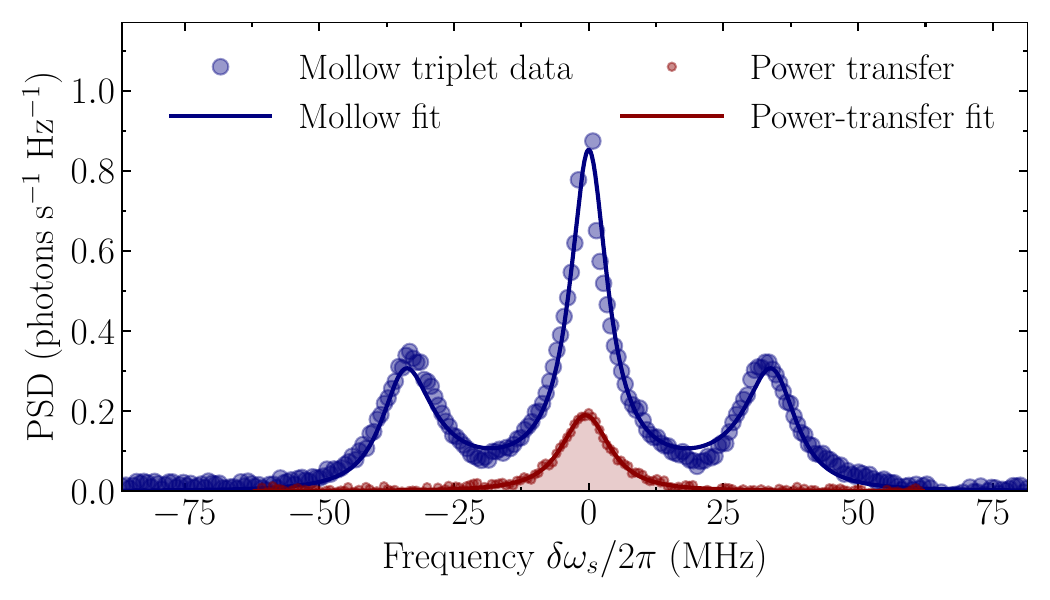}
    \caption{Calibration of the emitted power from the Mollow triplet and reference power-transport measurement. Blue points show the power spectral density of the resonance fluorescence from the driven symmetric transition measured in the resolved Mollow-triplet regime, with the solid blue line the fit to the standard two-level resonance-fluorescence spectrum obtained from the steady-state master equation and the quantum regression theorem. This fit provides the conversion between instrument units and emitted photon-flux spectral density at the output of the device. Red points show the differential emitted spectrum at the operating point used for the power-transport measurement, with the solid red line a Lorentzian fit. The area of this Lorentzian gives the heat current emitted into the symmetric waveguide and provides the absolute reference point used in the main TUR analysis.\label{S3}}
\end{figure}

The fit provides the conversion between instrument units and the emitted photon-flux spectral density at the device output before the amplification chain [Fig.~\ref{S3}]. This procedure fixes the overall gain of the analog and digital measurement chain, allowing us to report the power spectral density at the device output in physical units.

At the operating point used for the TUR experiment, we then measure the differential emitted spectrum between the ON and OFF configurations of the device. The resulting emission is well described by a Lorentzian line shape, which we denote by $L(\omega)$. The corresponding current we use as reference is then
\begin{equation}
J_{\rm reference} = \int_{-\infty}^{\infty} d\omega\, L(\omega),
\label{eq:SI_J_lorentzian}
\end{equation}
as indicated by the shaded region in  Fig~\ref{S3}.

We use this reference measurement to scale the subsequent time-domain power measurements. To do so, we measure the device at the same operating point and define the scaling factor 
\begin{equation}
    G = \frac{\tau J_{\rm reference}}{\langle \hat{Q}_\tau\rangle_{\rm on}-\langle \hat{Q}_\tau\rangle_{\rm off}}.
\end{equation}

This calibration then sets the scale for the first and second moments of the filtered integrated power by applying $G$ and $G^2$, respectively. We extract the moments from interleaved ON and OFF measurements performed for increasing drive rates. We obtain the mean heat current from Eq.~(\ref{eq:SI_J_onoff}), determine the measurement efficiency from the OFF-state data using Eq.~(\ref{eq:eff}), and extract the diffusion coefficient from the ON/OFF variance difference using Eq.~(\ref{eq:SI_boxcar_final}). Once $J$, $D$, and the cold-bath temperature are known, the TUR ratio follows directly from the entropy-production estimate discussed in the main text.

\subsection{Full counting statistics model}

The theory curves shown in the main text for the heat current, diffusion coefficient, and TUR ratio are obtained from a full counting statistics (FCS) calculation~\cite{landi_current_2024} for the same driven-dissipative molecular model used to interpret the experiment. We carry out the calculation in the three-state basis where the coherent drive acts on the antisymmetric transition and the emitted current of interest is the net photon flux through the symmetric output channel.

In this reduced description, the Hamiltonian in the rotating frame is written as
\begin{equation}
H = \hbar\delta_a\,|a\rangle\langle a|+\hbar\omega_s\,|s\rangle\langle s|+\frac{\hbar\Omega_a}{2}\left(|a\rangle\langle 0|+|0\rangle\langle a|\right),
\label{eq:SI_FCS_H}
\end{equation}
where $\delta_a$ is the detuning from the antisymmetric mode frequency, and $\Omega_a$ is the applied drive amplitude on the antisymmetric transition. We describe dissipation by jump operators that represent radiative exchange with the principal symmetric and antisymmetric baths, weaker auxiliary loss channels associated with the base-temperature environment, the noise-induced transfer channel discussed above, and residual pure dephasing.

A representative jump-operator set is
\begin{align}
J_1 &= \sqrt{\Gamma_s(n_s+1)}\,|0\rangle\langle s|, & J_2 &= \sqrt{\Gamma_s n_s}\,|s\rangle\langle 0|, \notag\\
J_3 &= \sqrt{\Gamma_{s,2}(n_s+1)}\,|0\rangle\langle s|, & J_4 &= \sqrt{\Gamma_{s,2} n_s}\,|s\rangle\langle 0|, \notag\\
J_5 &= \sqrt{\Gamma_{a}(n_a+1)}\,|0\rangle\langle a|, & J_6 &= \sqrt{\Gamma_{a} n_a}\,|a\rangle\langle 0|, \notag\\
J_7 &= \sqrt{\gamma_{\phi,a}^{\rm pure}/2}\,|a\rangle\langle a|, &
J_8 &= \sqrt{\gamma_{\phi,s}^{\rm pure}/2}\,|s\rangle\langle s|, \notag\\
J_9 &= \sqrt{\Gamma_{\phi}/2}\,\left(|s\rangle\langle a|+|a\rangle\langle s|\right),
\label{eq:SI_FCS_jops}
\end{align}
Here, $n_s$ and $n_a$ denote the thermal populations of the environment at the symmetric and antisymmetric mode frequencies, respectively, and are given by the Bose–Einstein distribution at temperature $T$. The rate $\Gamma_{s,2}$ denotes a weak background loss channel of the symmetric mode. This rate is much smaller than the dominant radiative decay into the symmetric waveguide and we take it to be comparable to the measured decay rate of the antisymmetric mode, $\Gamma_{a}/2\pi = 30$ kHz. Because this channel cannot be resolved independently, we evaluate the theory curves in Fig.~\ref{fig4} with $\Gamma_{s,2}/2\pi = 0$ kHz. The predicted minimum TUR ratio is then 1.73, compared with 1.75 for $\Gamma_{s,2}/2\pi = 30$ kHz, indicating that the result is only weakly affected by this uncertainty. We independently obtain the remaining parameters from spectroscopy and temperature-estimation measurements.

\begin{figure}[t]
    \includegraphics[width=8.5cm]{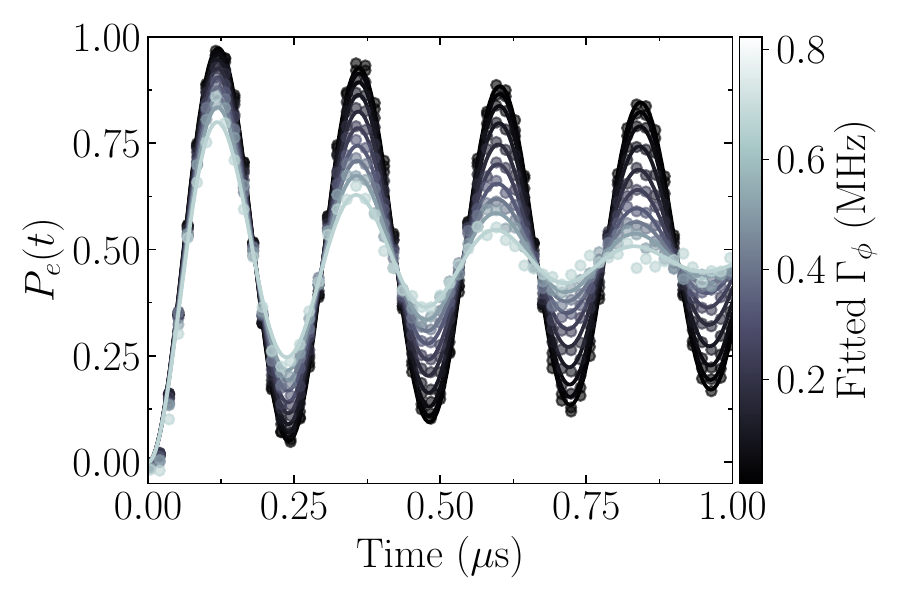}
    \caption{Rabi measurements of the antisymmetric mode for varying noise powers. Solid lines are fits to a Lindblad model used to extract the noise-induced transfer rate $\Gamma_{\Phi}$.\label{Rabi}}
\end{figure}

To compute the cumulants of the current into waveguide S, we attach a counting field $\chi$ to the jump operators associated with the transport channel of interest. This means dressing the principal emission and absorption operators with opposite counting phases. The tilted Liouvillian may then be written as
\begin{equation}
\mathcal{L}_{\chi}(\rho)= \frac{1}{i\hbar}[H,\rho] + \sum_j e^{-i c_j \chi} J_j\rho J_j^{\dagger} - \frac{1}{2}\sum_j\left\{J_j^{\dagger}J_j,\rho\right\},
\label{eq:SI_tilted_liouvillian}
\end{equation}
where $c_j=\pm1$ for jump operators $J_1$ and $J_2$ respectively and $c_j=0$ for the other jump operators. The long-time cumulant generating function is given by the eigenvalue $\lambda(\chi)$ of $\mathcal{L}_{\chi}$ with the largest real part. We obtain the first and second photon-current cumulants from derivatives at $\chi=0$,
\begin{equation}
J = \left. i\,\partial_{\chi}\lambda(\chi)\right|_{\chi=0},
\qquad
D = \left. -\partial_{\chi}^2\lambda(\chi)\right|_{\chi=0},
\label{eq:SI_FCS_cumulants}
\end{equation}
which we evaluate numerically by symmetric finite differences around $\chi=0$.

We obtain the heat current and diffusion coefficient plotted in the main text by multiplying these photon-current cumulants by the energy of the emitted photons at the symmetric transition. To evaluate the entropy production entering the theoretical TUR ratio, we also include heat exchanged through the antisymmetric radiative channel, described by the jump operators \(J_{5,6}\). The theory curves in Fig.~\ref{fig4} therefore account for heat flow through both the symmetric and antisymmetric dissipative channels. The flux noise is described by Hermitian jump operators $J_{7,8,9}$, equivalent to infinite-temperature or pure-dephasing noise, and therefore does not contribute to entropy production~\cite{landi_current_2024}. All calculations use the same calibrated frequencies, occupations, and decay rates that enter the spectroscopy and thermometry analysis.

\subsection{Independent extraction of the noise-induced transfer rate}
We extract the noise-induced mixing rate \(\Gamma_{\Phi}\) independently from time-domain Rabi measurements of the antisymmetric mode performed at increasing noise powers [see Fig.~\ref{Rabi}]. We fit the measured traces using the same Lindblad framework as in the transport FCS model, expressed here in a simple qutrit basis with the dephasing channel coupling the upper two levels. We first determine the baseline relaxation and dephasing rates from zero-noise data, and then extract the additional rate \(\Gamma_{\Phi}\) from the decay of the oscillations at finite noise power. This procedure calibrates \(\Gamma_{\Phi}\) independently of the transport and full-counting-statistics analyses, although the same parameter could equivalently be inferred by fitting the power spectral density or heat-flow data directly.

\subsection{Uncertainty estimation}

At each data point in Figure \ref{fig4}, a total of $5 \times 10^6$ averages are accumulated to construct the averaged moments, and the entire procedure is repeated at least 20 times. The uncertainty is written as $\sigma_{\mathrm{tot}}=\sqrt{\sigma_{\mathrm{samp}}^2+\sigma_G^2+\sigma_T^2}$, where $\sigma_{\mathrm{samp}}$ is the statistical uncertainty from the repeated measurements and account for most of the total uncertainty, $\sigma_G$ is the propagated uncertainty associated with the power-calibration factor $G$, and $\sigma_T$ is the propagated uncertainty from the cold-bath temperature estimation. The uncertainty in $G$ is inherited from the uncertainty in $J_{\mathrm{reference}}$, which in turn is set by the uncertainty of the scaling parameter obtained from fitting the Mollow-triplet calibration to the measured spectrum. To propagate $\sigma_G$ and $\sigma_T$, we recompute the full analysis chain: the fitted uncertainties in $J_{\mathrm{reference}}$ and in the cold-bath temperature are propagated through $G$, $\eta$, $J$, $D$, and finally $\mathcal{Q}$. The resulting contributions are then combined to obtain the uncertainty of $2\sigma_{\mathrm{tot}}$ main text.

A separate systematic uncertainty concerns dissipative channels not included in the entropy-production estimate. The noise-induced transfer channel does not contribute within our effective-reservoir model: the classical noise drives the \(|a\rangle\leftrightarrow|s\rangle\) transitions at equal rates, corresponding through detailed balance to \(T_\phi\rightarrow\infty\), so that its entropy-flow contribution \(-J_\phi/T_\phi\) vanishes even for finite \(J_\phi\). The energy exchanged in each such transition, \(\hbar 2g\), is also only \(2g/\omega_s\simeq0.07\) of the energy carried by an emitted symmetric photon. Additionally, the theoretically calculated entropy production already includes the measured dissipation of the antisymmetric mode through the jump operators \(J_{5,6}\) in Eq \ref{eq:SI_FCS_jops}. The remaining unresolved contribution is nonradiative decay of the symmetric mode. Taking its rate to be \(\Gamma_{s,2}/2\pi=30\)~kHz, equal to the measured antisymmetric decay rate and approximately \(0.4\%\) of the dominant symmetric radiative rate, increases the predicted minimum TUR ratio from \(1.73\) to \(1.75\). The predicted violation therefore remains robust to this additional dissipation channel.

\pagebreak
\widetext
\newpage
\begin{center}
\textbf{\large Supplementary Information for ``A quantum thermal machine surpassing the classical thermodynamic limit on precision''}
\end{center}
\setcounter{equation}{0}
\setcounter{figure}{0}
\setcounter{table}{0}
\setcounter{page}{1}
\makeatletter
\renewcommand{\theequation}{S\arabic{equation}}
\renewcommand{\thefigure}{S\arabic{figure}}
\renewcommand{\bibnumfmt}[1]{[S#1]}

\section{Experimental setup}

\begin{figure}[h]
    \includegraphics[width=18.5cm]{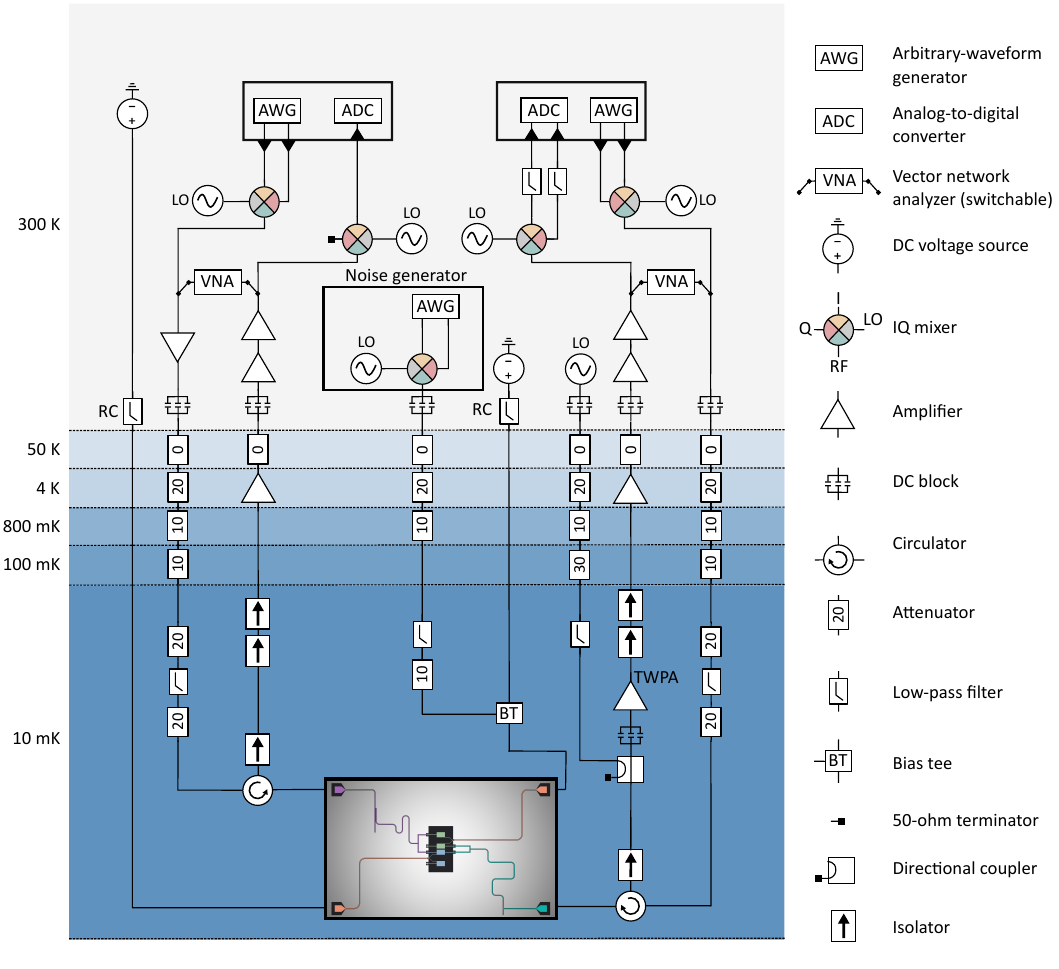}
    \caption{Experimental wiring diagram. The device is mounted at the mixing-chamber stage of the dilution refrigerator and accessed via symmetry-selective microwave lines addressing the symmetric and antisymmetric waveguide modes. Input signals are attenuated and thermalized at successive cryogenic stages, while output signals are routed through circulators, a TWPA, and additional cryogenic and room-temperature amplification. A vector network analyzer is used for continuous-wave reflection spectroscopy, and an IQ-based microwave transceiver is used for time-domain, spectral, and power measurements. Broadband flux noise centered near the molecular splitting $2g$ is injected through the flux line to induce noise-assisted transport.\label{S2} }
\end{figure}
The experimental setup for our study is depicted in Fig.~\ref{S2}. Our device is placed at the mixing chamber stage of a dilution refrigerator, maintaining a stable temperature of 10 mK. To isolate from external interference's, the device is housed within a copper sample holder which in turn is encased in a copper enclosure for electromagnetic wave shielding and a $\mu$-metal enclosure for protection against low-frequency magnetic fields.

The routing of input and output signals is managed by a microwave circulator, enabling a reflection measurement setup to both the symmetric and antisymmetric waveguide. For the input lines going to our device, we employ highly attenuated coaxial lines. The signals emanating from the device are then captured in output lines equipped with a High Electron Mobility Transistor (HEMT) amplifier at the 4 K stage. Additional amplifiers are used at room temperature (300K) to further amplify the signal. For the readout on the symmetric side, we also use a TWPA from Arctic Instruments.

For our measurements we physically toggle between (indicated by a switch in Fig.~\ref{S2}) a vector network analyzer (VNA) and a microwave transceiver where the latter is used in conjunction with IQ mixers. The VNA is used for continuous-wave reflection spectroscopy and the microwave transceiver for frequency-resolved measurements and power measurements. For the microwave transceiver we utilize a Quantum Machines OPX+ which is comprised of both arbitrary waveform generators (AWG) and analog-to-digital converters (ADC). 

To populate the flux line with white noise we utilize arbitrary waveform generators of the Keysight 3202A model to continuously synthesize voltage noise with a finite bandwidth of 50 MHz, upconverted to be centered around the energy gap $2g$ of the qubits.

\section{Input-output theory calculation of heat current and diffusion coefficient}

We derive here the expressions used in the main text to extract the mean heat current and its diffusion coefficient from continuous heterodyne measurements of the amplified output field. We start with a general input-output treatment for a filtered heterodyne signal and then specialise to the boxcar filter used in the experiment.

We begin from the input-output relation~\cite{Yurke1984, Gardiner1985,clerk_introduction_2010}
\begin{equation}
b_{\mathrm{o}}(t)=b_{\mathrm{i}}(t)+L(t),
\end{equation}
where $b_{\mathrm{i}}(t)$ and $b_{\mathrm{o}}(t)$ are the input and output fields, and $L(t)$ is the system operator coupling the device to the measured reservoir. In our case, this corresponds to the jump operator associated with the principal emission channel into the symmetric waveguide, $L = \sqrt{\Gamma_s(n+1)}\,|0\rangle\langle s| $, as defined in the main text.

The fields obey the usual commutation relations
\begin{align}
[b_{\mathrm{i}}(t),b_{\mathrm{i}}^{\dagger}(t^\prime)]=\delta(t-t^\prime),\qquad [b_{\mathrm{i}}(t),b_{\mathrm{i}}(t^\prime)]=0,  \notag\\
[b_{\mathrm{o}}(t),b_{\mathrm{o}}^{\dagger}(t^\prime)]=\delta(t-t^\prime), \qquad [b_{\mathrm{o}}(t),b_{\mathrm{o}}(t^\prime)]=0.
\end{align}
Throughout this section we assume vacuum input, $\langle b_{\mathrm{i}}^{\dagger}(t)b_{\mathrm{i}}(t^\prime)\rangle=0$.

After phase-insensitive linear amplification in the large-gain limit, the detected signal can be written as
\begin{equation}
S(t)=b_{\mathrm{o}}(t)+h^{\dagger}(t),
\end{equation}
where $h(t)$ is an auxiliary bosonic noise mode satisfying
\begin{equation}
[h(t),h^{\dagger}(t^\prime)]=\delta(t-t^\prime),\qquad [h(t),h(t^\prime)]=0.
\end{equation}
The added-noise mode is statistically independent of the output field and is assumed to have stationary occupation $n_h$, such that
\begin{equation}
\langle h^{\dagger}(t)h(t^\prime)\rangle=n_{\mathrm{h}}\,\delta(t-t^\prime).
\end{equation}
In this convention the measurement efficiency is
\begin{equation}
\eta=(1+n_{\mathrm{h}})^{-1}.
\end{equation}

To account for finite detection bandwidth, we filter the signal with a kernel $f(t)$,
\begin{equation}
\bar{S}(t)=\int_0^t dt^\prime\, f(t-t^\prime)S(t^\prime).
\end{equation}
We then define the integrated detected power over a time window $T$ as
\begin{equation}
\hat{Q}_T=\int_0^T dt\,\bar{S}^{\dagger}(t)\bar{S}(t),
\end{equation}
which corresponds to the total measured photon number in the time interval $[0,T]$.

For later use we introduce
\begin{equation}
\lambda(t)=\int_0^t dt_1\, |f(t_1)|^2,
\end{equation}
and the filtered field correlators
\begin{equation}
\bar{g}^{(1)}(t,t^\prime)=\int_0^t dt_1\int_0^{t^\prime} dt_2\, f^*(t-t_1)f(t^\prime-t_2)
\frac{\langle L^{\dagger}(t_1)L(t_2)\rangle}{\langle L^{\dagger}L\rangle},
\end{equation}
\begin{equation}
\bar{g}^{(2)}(t,t^\prime)=\int_0^{t^\prime} dt_1\int_0^t dt_2\int_0^t dt_3\int_0^{t^\prime} dt_4\,
f^*(t^\prime-t_1)f^*(t-t_2)f(t-t_3)f(t^\prime-t_4)
\frac{\langle L^{\dagger}(t_1)L^{\dagger}(t_2)L(t_3)L(t_4)\rangle}{\langle L^{\dagger}L\rangle^2},
\end{equation}
as well as the filter overlap
\begin{equation}
\Delta(t,t^\prime)=\int_0^{\min(t,t^\prime)} dt_1\, f^*(t-t_1)f(t^\prime-t_1).
\end{equation}

The mean integrated power then follows directly from the input-output relation together with $\langle b_{\mathrm{i}}^{\dagger}(t)b_{\mathrm{i}}(t^\prime)\rangle=0$ and $\langle h(t)h^{\dagger}(t^\prime)\rangle=(1+n_{\mathrm{h}})\delta(t-t^\prime)$:
\begin{align}
\langle \hat{Q}_T\rangle
&=\int_0^T dt\int_0^t dt_1\int_0^t dt_2\, f^*(t-t_1)f(t-t_2)
\langle b_{\mathrm{o}}^{\dagger}(t_1)b_{\mathrm{o}}(t_2)+h(t_1)h^{\dagger}(t_2)\rangle \notag\\
&=\int_0^T dt\left[J\bar{g}^{(1)}(t,t)+\frac{\lambda(t)}{\eta}\right],
\label{eq:SI_Q_general}
\end{align}
where
\begin{equation}
J=\langle L^{\dagger}L\rangle
\label{eq:SI_J_def}
\end{equation}
is the steady state emitted photon current. 

We define the first and second-order correlation functions of the emitted field in terms of the jump operator $L$ as
\begin{equation}
    g^{(1)}(\tau) = \frac{\langle L^\dagger(t) L(t+\tau)  \rangle}{J}, \qquad  g^{(2)}(\tau) = \frac{\langle L^\dagger(t) L^\dagger(t+\tau) L(t+\tau) L(t)  \rangle}{J^2},
    \label{eq:SI_g1_g2_def}
\end{equation}
where in steady state, both functions depend only on the time delay $\tau$.

In the experimentally relevant limit where the filter memory time $\Lambda^{-1}$ is short compared with the correlation time of $g^{(1)}(\tau)$, and $\Lambda T\gg 1$, we approximate $\bar{g}^{(1)}(t,t^\prime)\to g^{(1)}(t-t^\prime)$, resulting in $\bar{g}^{(1)}(t,t)\simeq g^{(1)}(0)=1$, and $\lambda(t)\to \Lambda$, with
\begin{equation}
\Lambda=\int_0^{\infty}dt\, |f(t)|^2.
\end{equation}
This yields
\begin{equation}
\frac{\langle \hat{Q}_T\rangle}{T}=J+\eta^{-1}\Lambda.
\label{eq:SI_mean_current_general}
\end{equation}
This intuitive result can be interpreted as the total measured power being the photon current emitted from the system in addition to the total recorded noise given by the measurement bandwidth times the added noise per mode. Subtracting the OFF-state data removes the bandwidth-dependent noise and yields the current estimator used in the main text,
\begin{equation}
J=\frac{1}{T}\left[\langle \hat{Q}_T\rangle_{\rm on}-\langle \hat{Q}_T\rangle_{\rm off}\right].
\label{eq:SI_J_onoff}
\end{equation}

To access fluctuations we next evaluate the second moment of $\hat{Q}_T$. In the large-gain limit, the amplified signal commutes with itself at different times, allowing us to write
\begin{align}
\langle \hat{Q}_T^2\rangle
&=2\int_0^T dt\int_0^t dt^\prime\,\langle \bar{S}^{\dagger}(t^\prime)\bar{S}^{\dagger}(t)\bar{S}(t)\bar{S}(t^\prime)\rangle \notag\\
&=2\int_0^T dt\int_0^t dt^\prime\Bigg[J^2\bar{g}^{(2)}(t,t^\prime)+\eta^{-2}\lambda(t)\lambda(t^\prime)+\eta^{-2}|\Delta(t,t^\prime)|^2 \notag\\
&\qquad\qquad +J\eta^{-1}\lambda(t)\bar{g}^{(1)}(t^\prime,t^\prime)+J\eta^{-1}\lambda(t^\prime)\bar{g}^{(1)}(t,t) \notag\\
&\qquad\qquad +2J\eta^{-1}\operatorname{Re}\!\left[\Delta(t,t^\prime)\bar{g}^{(1)}(t^\prime,t)\right]\Bigg].
\label{eq:SI_Q2_general}
\end{align}
Here we have assumed Gaussian statistics for the added-noise mode, so that Wick's theorem applies when reducing fourth-order noise correlators.

Under the same large-bandwidth assumptions used above, we approximate  $\bar{g}^{(1)}(t,t^\prime)\to g^{(1)}(t-t^\prime)$, $\bar{g}^{(2)}(t,t^\prime)\to g^{(2)}(t-t^\prime)$, and $\Delta(t,t^\prime)\to \delta(t-t^\prime)$. For an exponential low-pass filter,
\begin{equation}
f(t)=2\Lambda e^{-2\Lambda t},
\end{equation}
this yields
\begin{equation}
\frac{{\rm Var}[\hat{Q}_T]}{T}=\frac{\Lambda}{2\eta^2}+\frac{2J}{\eta}+\frac{2J^2}{T}\int_0^T dt\int_0^t dt^\prime\,[g^{(2)}(t-t^\prime)-1].
\end{equation}
For $T$ much longer than the correlation time of $g^{(2)}$, this expression defines the diffusion coefficient \cite{landi_current_2024},
\begin{equation}
D=J+2J^2\int_0^{\infty} d\tau\,[g^{(2)}(\tau)-1],
\label{eq:SI_D_def}
\end{equation}
and hence we get the general relation quoted in the main text,
\begin{equation}
\frac{1}{T}\left({\rm Var}[\hat{Q}_T]_{\rm on}-{\rm Var}[\hat{Q}_T]_{\rm off}\right)=D+\frac{2-\eta}{\eta}J.
\label{eq:SI_D_onoff_general}
\end{equation}

Equations~(\ref{eq:SI_J_def}) and (\ref{eq:SI_D_def}) correspond to the first and second cumulants of the \emph{emitted current} obtained from the full counting statistics model described in the main text. This formulation assumes that we only count the photons emitted into the measured channel, as described by the jump operator $L$. So the resulting statistics describe the emitted current rather than the net current given by the difference between the emitted and absorbed current. However, in the low-temperature regime considered in our work, absorption processes associated with the cold symmetric waveguide are negligible and do not contribute significantly.

Note that, for a bosonic input field $b_{\mathrm{i}}(t)$ with finite thermal occupation $\langle b_{\mathrm{i}}(t) b_{\mathrm{i}}^\dagger(t^\prime)\rangle = n_{\mathrm{i}}$, the above results remain valid upon redefining the efficiency as $\eta = 1/(1+n_{\mathrm{h}} +n_{\mathrm{i}})$. 

We now adapt this general framework to the boxcar filter used in the experiment.

\subsection{Filter function and measurement-efficiency corrections}

In the experiment, the filtered integrated power is not obtained with an exponential low-pass kernel but with a broad boxcar filter. We therefore repeat the evaluation above for the filter actually used in the data processing.

For a boxcar filter of width $\Lambda^{-1}$,
\begin{equation}
f(t)=\Lambda\left[\Theta(t)-\Theta(t-\Lambda^{-1})\right],
\end{equation}
so that
\begin{equation}
\lambda(t)=\Lambda^2\left[t-(t-\Lambda^{-1})\Theta(t-\Lambda^{-1})\right].
\end{equation}
Substituting this into Eq.~(\ref{eq:SI_Q_general}) gives
\begin{equation}
\langle \hat{Q}_T\rangle=JT+\frac{\Lambda T}{\eta}-\frac{1}{2\eta},
\label{q1}
\end{equation}
or equivalently
\begin{equation}
\frac{\langle \hat{Q}_T\rangle}{T}=J+\frac{\Lambda}{\eta}-\frac{1}{2\eta T}.
\label{eq:SI_boxcar_mean_exact}
\end{equation}
Thus the boxcar filter produces a finite-$T$ correction of order $(\Lambda T)^{-1}$, while the ON/OFF current estimator remains unchanged:
\begin{equation}
J=\frac{1}{T}\left[\langle \hat{Q}_T\rangle_{\rm on}-\langle \hat{Q}_T\rangle_{\rm off}\right].
\end{equation}

The corresponding second moment is obtained by inserting the boxcar filter into Eq.~(\ref{eq:SI_Q2_general}). Under the same assumptions that the filter memory time is short compared with the decay times of $g^{(1)}$ and $g^{(2)}$, one finds
\begin{equation}
\frac{{\rm Var}[\hat{Q}_T]}{T}=D-J+\frac{2J}{\eta}-\frac{2J}{3\eta\Lambda T}+\frac{2\Lambda}{3\eta^2}-\frac{1}{2T\eta^2},
\label{eq:SI_boxcar_var_exact}
\end{equation}
with $D$ defined in Eq.~(\ref{eq:SI_D_def}). The corresponding ON/OFF subtraction yields
\begin{equation}
\frac{1}{T}\left({\rm Var}[\hat{Q}_T]_{\rm on}-{\rm Var}[\hat{Q}_T]_{\rm off}\right)
=D+J\left(\frac{2-\eta-\epsilon}{\eta}\right),
\end{equation}
where the finite-window correction scales as $\epsilon\sim (\Lambda T)^{-1}$. In the long-time limit $\Lambda T\gg 1$, this correction vanishes and we recover
\begin{equation}
\frac{1}{T}\left({\rm Var}[\hat{Q}_T]_{\rm on}-{\rm Var}[\hat{Q}_T]_{\rm off}\right)=D+\frac{2-\eta}{\eta}J.
\label{eq:SI_boxcar_final}
\end{equation}

Eq.~(\ref{eq:SI_boxcar_final}) is the form used in the analysis of the experimental data and directly underlies the diffusion-coefficient extraction equation given in the main text. Thus, although the derivation is simplest for an exponential kernel, the boxcar filter used in the experiment leads to the same asymptotic relation between the ON/OFF power variance, the diffusion coefficient, and the excess-noise correction, up to subleading corrections in $(\Lambda T)^{-1}$.

From Eq. (\ref{eq:SI_boxcar_mean_exact}) and (\ref{eq:SI_boxcar_var_exact}) one also obtains a convenient estimate of the measurement efficiency directly from the OFF-state moments, 
\begin{equation}
\eta \approx \frac{2\langle \hat{Q}_T\rangle_{\rm off}}{3 {\rm Var}[\hat{Q}_T]_{\rm off}}.
\label{eq:eff}
\end{equation}

\subsection{Four-level model for cold-bath thermometry}
The retained basis that describes the lowest four states of the hybridized two-transmon molecule, are $|0\rangle$, the antisymmetric single-excitation state $|a\rangle$, the symmetric single-excitation state $|s\rangle$, and the lowest doubly excited state $|2+\rangle_L$ with a $|0\rangle\leftrightarrow|2+\rangle_L$ transition frequency at $\omega_2 / 2\pi = 11.69$ GHz. In this basis, the operators driving the symmetric and antisymmetric transitions are written as
\begin{equation}
b_S=
\begin{pmatrix}
0&0&1&0\\
0&0&0&0\\
0&0&0&\xi\\
0&0&0&0
\end{pmatrix},
\qquad
b_A=
\begin{pmatrix}
0&1&0&0\\
0&0&0&\sqrt{2}\\
0&0&0&0\\
0&0&0&0
\end{pmatrix},
\end{equation}
where the parameter $\xi$ is the ratio of dipole matrix elements, nominally set to $\sqrt{2}$ for the second state transition in a transmon, but is here left as a fitting parameter due to frequency-dependent coupling to the transmission line through the filter resonator.

The spectroscopy tone driving the system at the frequency $\omega_d$ is applied through the symmetric waveguide. In the rotating frame, with the third and fourth basis states rotating at $\omega_d$ and $2\omega_d$, respectively, the Hamiltonian becomes
\begin{equation}
H=
\hbar \begin{pmatrix}
0 & 0 & \Omega_1/2 & 0 \\
0 & \omega_A & 0 & 0 \\
\Omega_1/2 & 0 & \omega_S-\omega_d & \,\Omega_2/\sqrt{2} \\
0 & 0 & \,\Omega_2/\sqrt{2} & \omega_{2}-2\omega_d
\end{pmatrix},
\end{equation}
where $\Omega_{1,2}$ is the applied drive amplitude for the two different symmetric transitions, which can be different due to the frequency dependent coupling through the filter resonator.

The dynamics of the system is well described by a Lindblad master equation with both symmetric and antisymmetric radiative channels included,
\begin{align}
\dot\rho ={}& \frac{1}{i\hbar}[H,\rho] \notag\\
&+\Gamma_s(n+1)\mathcal{D}[b_S](\rho)+\Gamma_s n\,\mathcal{D}[b_S^{\dagger}](\rho) \notag\\
&+\Gamma_a(n+1)\mathcal{D}[b_A](\rho)+\Gamma_a n\,\mathcal{D}[b_A^{\dagger}](\rho) \notag\\
&+\gamma_{\phi s}\mathcal{D}[b_S^{\dagger}b_S](\rho)+\gamma_{\phi a}\mathcal{D}[b_A^{\dagger}b_A](\rho).
\end{align}
from which the steady state density matrix $\rho_{\mathrm{ss}}$ can be obtained. Here, $n$ is the thermal occupation as seen by both the symmetric and antisymmetric modes and $\mathcal{D}[A](\rho)=A\rho A^\dagger-\tfrac{1}{2}A^\dagger A\rho-\tfrac{1}{2}\rho A^\dagger A$. For each probe frequency $\omega_d$, the coherent symmetric response is 
\begin{equation}
\langle b_S\rangle = \mathrm{Tr}(b_S\rho_{\mathrm{ss}}),
\end{equation}
which is then used to reconstruct the reflection coefficient obtained from standard input-output theory
\begin{equation}
r(\omega_d)=1-i\,2\Gamma_s\frac{\langle b_S\rangle}{\Omega}.
\label{reflEQ}
\end{equation}
To calibrate the elevated-temperature occupation, we probe only the second transition and fit the data with $\Omega_{1}=\Omega_{2}$, leaving the thermal occupation parameter $n$, the drive rate, and $\xi$ as free parameters. All remaining parameters are fixed independently from spectroscopy of the first symmetric and antisymmetric transitions. 

\subsection{Mollow-triplet power calibration}

We model the resonance fluorescence used for the absolute power calibration as that of a driven two-level system. In the rotating frame, the Hamiltonian is

\begin{equation}
H=\hbar\delta\,\sigma_+\sigma_-+\frac{\hbar\Omega}{2}(\sigma_++\sigma_-),
\end{equation}
and by the Liouvillian
\begin{align}
\mathcal{L}(\rho)={}&\frac{1}{i\hbar}[H,\rho]
+\Gamma n\,\mathcal{D}[\sigma_+](\rho)
+\Gamma(n+1)\,\mathcal{D}[\sigma_-](\rho) \notag \\
&+\frac{\gamma_\phi}{2}\,\mathcal{D}[\sigma_z](\rho).
\label{eq:SI_mollow_liouvillian}
\end{align}
Here $\Gamma$ is the radiative decay rate into the measured channel, $\gamma_\phi$ is the pure-dephasing rate, and $n$ is the thermal occupation of the bath mode at the transition frequency.

We solve for the steady state density matrix $\rho_{\mathrm{ss}}$ and evaluate the incoherent resonance-fluorescence spectrum through the resolvent form of the quantum regression theorem,
\begin{equation}
S(\omega)=\frac{\Gamma}{\pi}\operatorname{Re}\left[\operatorname{Tr}\left(\sigma_{+}(i \omega \mathbb{I}-\mathcal{L})^{-1}\left[\sigma_{-} \rho_{\mathrm{ss}}-\left\langle\sigma_{-}\right\rangle_{\mathrm{ss}} \rho_{\mathrm{ss}}\right]\right)\right]
\label{eq:SI_mollow_spectrum_raw_resolvent}
\end{equation}
which is equivalently the one-sided Fourier transform of the connected correlation function
\begin{equation}
S(\omega)=\frac{\Gamma}{\pi}\mathrm{Re}\!\int_0^{\infty} d\tau\, e^{-i\omega\tau}
\left[\langle \sigma_+(\tau)\sigma_-(0)\rangle-\left|\langle \sigma_-\rangle\right|^2\right].
\label{eq:SI_mollow_spectrum_raw}
\end{equation}
Subtracting $|\langle\sigma_-\rangle|^2$ removes the coherent elastic peak, so that only the incoherent Mollow-triplet contribution is fitted.

Fitting the measured Mollow-triplet spectrum to Eq.~(\ref{eq:SI_mollow_spectrum_raw_resolvent}) provides the conversion between instrument units and emitted power spectral density at the device output before the amplification chain (see result in Fig 6 in the main text). This procedure fixes the overall gain factor of our analog and digital measurement chain, allowing us to report the power spectral density at the device output in absolute units of photons per second per unit angular-frequency bandwidth.

\end{document}